\global\def\draftcontrol{0}

   \def\versionno{ New Directions }

\catcode`\@=11

\expandafter\ifx\csname draftcontrol\endcsname\relax\global\def\draftcontrol{0}
\fi

{\count255=\time\divide\count255 by 60
\xdef\hourmin{\number\count255}
\multiply\count255 by-60\advance\count255 by\time
\xdef\hourmin{\hourmin:\ifnum\count255<10 0\fi\the\count255}}
\def\draftdate{\number\month/\number\day/\number\year\ \ \ \hourmin }

\newcommand\makepapertitle{\par

  \begingroup
    \renewcommand\thefootnote{\@fnsymbol\c@footnote}%
    \def\@makefnmark{\rlap{\@textsuperscript{\normalfont\@thefnmark}}}%
    \long\def\@makefntext##1{\parindent 1em\noindent
            \hb@xt@1.8em{%
                \hss\@textsuperscript{\normalfont\@thefnmark}}##1}%
     \newpage
     \global\@topnum\z@   
     \@makepapertitle
     \thispagestyle{empty}\@thanks
  \endgroup
  \setcounter{footnote}{0}%
  \global\let\thanks\relax
  \global\let\makepapertitle\relax
  \global\let\@makepapertitle\relax
  \global\let\@thanks\@empty
  \global\let\@author\@empty
  \global\let\@date\@empty
  \global\let\@title\@empty
  \global\let\title\relax
  \global\let\author\relax
  \global\let\date\relax
  \global\let\and\relax
  \def\version{\let\version\@version\@gobble}
}
\def\@makepapertitle{%
  \newpage
   \ifnum\draftcontrol=1 {}
   \version\versionno
   \vskip 5em%
   \else
   \hfill\hbox to 4cm {\parbox{5cm}{\@pubnum}\hss}%
   \vskip 5em%
   \fi
   \begin{center}%
   \let \footnote \thanks
      {\hskip -0\textwidth \hbox to 1\textwidth%
        {\centerline{\Large\bf{\noindent\@title}}}}%
     \vskip 2em%
     {\normalsize
       \lineskip .5em%
       \begin{tabular}[t]{c}%
         \@author
       \end{tabular}\par}%
     \vskip 1em%
     {\@bstract}%
     \end{center}%
     \vfill
     \@date%
     \vskip 1.5em%
     \noindent
     \rule{12em}{.02em}\par\noindent
     \@email%
   \par
}

\gdef\@pubnum{}
\def\pubnum#1{%
  \gdef\@pubnum{#1}}

\gdef\@bstract{}
\def\Abstract#1{%
  \gdef\@bstract{%
   \parbox{\textwidth-0pc}{%
   \centerline{\bf Abstract}\penalty1000
   \noindent
   \renewcommand\baselinestretch{1.0}
   {#1}}}
}

\gdef\@email{}
\def\email#1{%
   \gdef\@email{%
   {\small Email: {\tt #1}}}
}

\def\ps@paper{\let\@mkboth\@gobbletwo%
     \ifnum\draftcontrol=1
        \def\@oddfoot{\hbox to \textwidth{\tiny \versionno \hfil\tiny\draftdate}%
        \hskip -\textwidth \hbox to \textwidth{\hfil\rm\thepage\hfil}}%
     \else\def\@oddfoot{\hbox to \textwidth{\hfil\rm\thepage\hfil}}
     \fi
     \let\@evenfoot\@oddfoot
}

\def\body{\clearpage
          \pagestyle{paper}
        }
\newenvironment{acknowledgments}{%
\vskip 3.25ex
\addcontentsline{toc}{section}{Acknowledgments}
\noindent {\bf Acknowledgments}
}

\def\@version#1{\ifnum\draftcontrol=1
\typeout{}\typeout{#1}\typeout{}
\vskip3mm\centerline{\hbox{\fbox{\normalsize{\tt DRAFT -- #1 -- }
                   {\draftdate}}}}\vskip3mm
\fi}
\let\version\@version
\long\def\eqlabel#1{\ifnum\draftcontrol=1
                    \tag@false  
                    \tag*{(\theequation) \hbox to -0.2cm{\hspace{0cm}\small{#1}\hss}}
                    \refstepcounter{equation}
                    \edef\@currentlabel{\theequation}
                    \ltx@label{#1}          
                    \else
                    \label{#1}
                    \fi
                    }
\let\st@bibitem\@bibitem
\let\st@lbibitem\@lbibitem
\ifnum\draftcontrol=1
  \def\@bibitem#1{%
    \st@bibitem{#1}\a@@label{#1}\ignorespaces}
  \def\@lbibitem[#1]#2{%
    \st@lbibitem[#1]{#2}\a@@label{#2}\ignorespaces}
  \def\a@@label#1{%
    \gdef\a@lab{\smash{\normalfont\small#1}}
    \ifvmode
      \if@inlabel
        \global\setbox\@labels\hbox{%
          \llap{\a@lab\let\a@lab\relax
                \kern\@totalleftmargin\kern\marginparsep}%
          \box\@labels}%
      \fi
    \fi}
\fi

\documentclass[12pt,letterpaper]{article}
\usepackage{graphicx}
\usepackage{amsmath,bm,amsfonts,amssymb,array,calc,amsthm,rotating}
\usepackage{epsfig}
\usepackage{graphicx}
\usepackage{color}
\usepackage[colorlinks=false]{hyperref}

\renewcommand\baselinestretch{1.25}
\renewcommand\section{\@startsection {section}{1}{\z@}%
                                   {-3.5ex \@plus -1ex \@minus -.2ex}%
                                   {2.3ex \@plus.2ex}%
                                   {\normalfont\large\bfseries}}
\renewcommand\subsection{\@startsection{subsection}{2}{\z@}%
                                   {-3.25ex\@plus -1ex \@minus -.2ex}%
                                   {1.5ex \@plus .2ex}%
                                   {\normalfont\normalsize\bfseries}}
\renewcommand\subsubsection{\@startsection{subsubsection}{3}{\z@}%
                                   {-3.25ex\@plus -1ex \@minus -.2ex}%
                                   {1.5ex \@plus .2ex}%
                                   {\normalfont\normalsize\it}}
\renewcommand\paragraph{\@startsection{paragraph}{4}{\z@}%
                                   {-3.25ex\@plus -1ex \@minus -.2ex}%
                                   {1.5ex \@plus .2ex}%
                                   {\normalfont\normalsize\bf}}
\renewcommand\subparagraph{\@startsection{subparagraph}{5}{\z@}%
                                   {-1.25ex\@plus -1ex \@minus -.2ex}%
                                   {0ex \@plus .2ex}%
                                   {\normalfont\normalsize\it}}

\numberwithin{equation}{section}

\long\def\@makecaption#1#2{%
  \vskip\abovecaptionskip
  \sbox\@tempboxa{{\bf #1:} #2}%
  \ifdim \wd\@tempboxa >\hsize
    {\small\bf #1:} {\small #2}\par
  \else
    \global \@minipagefalse
    \hb@xt@\hsize{\hfil\box\@tempboxa\hfil}%
  \fi
  \vskip\belowcaptionskip}

\renewcommand*\l@section[2]{%
  \ifnum \c@tocdepth >\z@
    \addpenalty\@secpenalty
    \addvspace{.5em \@plus\p@}%
    \setlength\@tempdima{1.5em}%
    \begingroup
      \parindent \z@ \rightskip \@pnumwidth
      \parfillskip -\@pnumwidth
      \leavevmode \bfseries
      \advance\leftskip\@tempdima
      \hskip -\leftskip
      #1\nobreak\hfil \nobreak\hb@xt@\@pnumwidth{\hss #2}\par
    \endgroup
  \fi}
\renewcommand*\l@subsection{\addvspace{.0em \@plus\p@}\@dottedtocline{2}{1.5em}{2.3em}}
\renewcommand*\l@subsubsection{\addvspace{-.2em \@plus\p@}\@dottedtocline{3}{3.8em}{3.2em}}

\def\hepth#1{\href{http://xxx.arxiv.org/abs/hep-th/#1}{{arXiv:hep-th/#1}}}

\def\arxiv#1#2{\href{http://xxx.arxiv.org/abs/#1}{{arXiv:#1 [#2]}}}

\definecolor{refcol}{rgb}{0.2,0.2,0.8}
\definecolor{eqcol}{rgb}{.6,0,0}
\definecolor{purple}{cmyk}{0,1,0,0}

\gdef\@citecolor{refcol}
\gdef\@linkcolor{eqcol}
\def\colorlinkspurple{\gdef\@urlcolor{purple}}
\def\colorlinksblue{\gdef\@urlcolor{blue}}
\def\colorlinksred{\gdef\@urlcolor{red}}

\def\revise#1       {\raisebox{-0em}{\rule{3pt}{1em}}%
                     \marginpar{\raisebox{.5em}{\vrule width3pt\
                     \vrule width0pt height 0pt depth0.5em
                     \hbox to 0cm{\hspace{0cm}{%
                     \parbox[t]{4em}{\raggedright\footnotesize{#1}}}\hss}}}}

\def\caln         {{\cal N}}

\def\zet          {{\mathbb Z}}

\def\ee           {{\it e}}
\def\ii           {{\it i}}

\def\sqr#1#2{{\vcenter{\vbox{\hrule height.#2pt
 \hbox{\vrule width.#2pt height#1pt \kern#1pt
 \vrule width.#2pt}\hrule height.#2pt}}}}

  \def\a{\alpha}
  \def\b{\beta}

\catcode`\@=12

\begin{document}


\title{Runaway in the Landscape}

\pubnum{%
\arxiv{0706.0514}{hep-th}}
\date{June 2007}

\author{Katrin Becker$^{a}$, Melanie Becker$^{a,b}$
and Johannes Walcher$^{c}$ \ \\[0.4cm]
\it $^a$ Texas A \& M University, College Station, TX 77843, USA
\\[0.2cm]
\it $^b$ Harvard University, Cambridge, MA 02138, USA \\[0.2cm]
\it $^c$ Institute for Advanced Study,
\it Princeton, NJ 08540, USA}

\Abstract{We consider flux compactifications of type IIB string
theory on the mirror of a rigid Calabi-Yau. In special cases, these
models are dual to the type IIA flux vacua with runaway direction in
flux space. We show that new weak coupling AdS solutions can be
found for large complex structure, while Minkowski solutions with
all moduli stabilized are confined to be at strong coupling. The
existence of these solutions, as found in a previous work, is
nevertheless guaranteed by a non-renormalization theorem of the type
IIB flux superpotential. Based on our results, we are led to the
conjecture that supersymmetric runaway directions in flux space are
always accompanied by a spectrum of moduli masses reaching down to
the AdS scale. This could be violated in a non-supersymmetric
situation.}

\email{kbecker@physics.tamu.edu, mbecker@physics.tamu.edu, walcher@ias.edu}


\makepapertitle

\body

\version\versionno

\vskip 1em



\section{Introduction}

One of the most important lessons we learned about flux
compactifications in the past couple of years is that moduli
fields of the internal geometry can be stabilized in these
compactifications (for a review and an extensive list of references
see \cite{Grana:2005jc},\cite{dk:2006}). In a vast majority of the
compactifications the addition of non-perturbative effects coming
from gluino condensation or wrapped branes is needed in order to
stabilize all the moduli. This is not totally unexpected \cite{dise}
if one insists on working with a small parameter controlling a
semi-classical expansion.

It therefore came as a surprise when an explicit model with non-zero
masses for all geometric moduli fields at the classical level ({\it
i.e.} in terms of fluxes only) appeared in \cite{wati} in the
context of the massive type IIA theory. One of the most interesting
aspects of this proposal is that it exhibits a direction in the
space of flux configurations on the internal manifold which is not
constrained by tadpole cancellation while maintaining $\caln=1$
supersymmetry. Along this direction, one gains back a small
parameter, the inverse flux number, which can be used for a
perturbative expansion.

One may wonder if by going to the large flux limit one isn't
throwing out the baby with the bathwater. Namely, it has been
emphasized that the masses of the geometric moduli appearing in
\cite{wati} are actually proportional to the cosmological constant
of the four-dimensional anti de Sitter (AdS) space. In other words,
their Compton wavelengths are comparable to the AdS radius. So just
as ``tachyons'' with negative mass squared above the
Breitenlohner-Freedman bound, those moduli can be considered as
effectively massless for all practical purposes. Note that this does
not stand in the way of the fact that the vacua of \cite{wati} are
indeed four-dimensional because the Kaluza-Klein scale is
parametrically larger that the inverse AdS radius.

The question arises whether this sort of behavior of moduli masses
will bear out in any runaway direction in the landscape in which one
is gaining parametric control over the expansion by dialing some
fluxes to be large. In this paper, we will find evidence that the
answer to this question is in the affirmative. We will address this
question by extending the type IIB models recently proposed in
\cite{BVW}. Similar in spirit as in \cite{wati} the models proposed
in \cite{BVW} have all moduli stabilized in terms of fluxes only,
but in distinction to those of \cite{wati}, were constructed without
any small parameter at all. Control is provided by a powerful
non-renormalization theorem for the flux superpotential.

Our first aim is to deform the models of \cite{BVW} and show that
they exhibit a runaway direction in flux space similar to
\cite{wati}. In the context of this model it will then be shown in
section 4 that parametric control over the expansion is indeed
closely related to masses of the size of the AdS scale.

The advantage of working in type IIB is that one has technical
control over a larger variety of fluxes than in type IIA. This
greater amount of freedom in dialing the fluxes might lead to the
expectation that the masses of moduli are not necessarily linked to
the AdS scale. We will not be able to realize this expectation, and
although we will come relatively close, we view our failure as
evidence that it might simply not be possible to do so. In fact, we
conjecture that {\it for any sequence of supersymmetric weakly
coupled string vacua reaching out to the boundary of moduli space
the masses of some of the moduli fields are of the order of the
cosmological constant.}\footnote{It seems likely that such a generic
constraint on AdS vacua could be effectively studied using general
properties of the putative dual conformal field theories. Such an
approach was advocated by many people and was used concretely in
\cite{Ooguritwo} in constraining the mass gap for AdS$_3$ vacua
using the modularity of 2d CFT partition function.} Moduli fields
emerging in compactifications of string theory to AdS or Minkowski
space can be stabilized at best at strong coupling and could only be
described explicitly if a weakly coupled dual conformal field theory
with a large mass gap can be constructed.

Our conjecture is somewhat reminiscent of one of the general swampland
conjectures of \cite{Ooguri} according to which boundary regions of
moduli space are always signaled by the appearance of new light
fields in the spectrum. This might help to further delineate
features of the swampland \cite{swamp}, and also gives support to
the conjecture that the number of string vacua passing the most
basic of physical cuts is finite \cite {Ooguri}, \cite{finite}.

It is expected that the situation changes after supersymmetry is
broken (and the vacua are uplifted to de Sitter (dS) space), because
then moduli fields may acquire masses that in principle could be
much larger than the cosmological constant. Flux compactifications
offer several mechanisms for supersymmetry breaking. In the type IIB
context, the main mechanism that uplifts the AdS vacua to dS space
is the one proposed in \cite{Kachru} by using anti-branes. This is
close to the local models of \cite{kpv}. In
\cite{Aganagic},\cite{Heckman} it was proposed that a system of
branes and anti-branes on the resolved conifold is holographically
dual to a flux configuration with appropriate sign flips on the
deformed conifold. The advantage of the dual formulation is that
supersymmetry can be broken spontaneously to ${\cal N}=0$ in terms
of only fluxes. One may wonder if supersymmetry can be broken
spontaneously for the compact models considered herein. In the type
IIA context, such a mechanism was considered in \cite{vandoren}.
(See \cite{Kallosh} for a general discussion of uplifting in IIA.)
The type IIB version, on which we shall elaborate in Section 5, has
two ingredients (which are mirror dual to those considered in
\cite{vandoren}). To begin with, one turns on only fluxes of RR
type, which by themselves break supersymmetry {\it spontaneously},
but lead to a runaway potential for the dilaton. This runaway can
then be stabilized similarly to \cite{Kachru} in the presence of
suitable non-perturbative effects, and lead to a vacuum with a
positive cosmological constant! We will also show that masses of
moduli can become much larger than the scale of the cosmological
constant in this type of vacua.

A byproduct of our analysis is that the models first constructed in
\cite{wati} in type IIA in fact admit a {\it mirror dual}
description in terms of a more familiar construction in type IIB.
Namely, all the fluxes considered in \cite{wati} (in their explicit
examples, not necessarily the general case) in fact transform under
mirror symmetry into ordinary RR and NS-NS three-form fluxes. The
mirror transform can be done using the SYZ approach by taking into
account that the 3 T-dualities are done along the directions in
which there is no $H_{NS}$ flux. This claim might be surprising if
one recalls that in the usual type IIB construction, supersymmetric
flux configurations have an imaginary self-dual (ISD) property and
therefore lead to a positive definite contribution to the D3-brane
tadpole. As a consequence, there should not be any supersymmetric
runaway direction in flux space which is not constrained by the
tadpole. The reason for the unexpected existence of a runaway
direction in flux space is an unusual prefactor in the K\"ahler
potential for the dilaton. This prefactor is peculiar to the {\it
non-geometric models} we consider here, and is predicted by mirror
symmetry as we explain in some detail. Supersymmetric fluxes in the
non-geometric type IIB models are not constrained to be ISD.

This paper is organized as follows. In section 2 we review some of
the relevant formulas of the type IIB model constructed in
\cite{BVW} that will be needed in the later sections. Due to its
importance in connection to the existence of strong coupling
solutions, we will elaborate the non-renormalization theorem
presented in \cite{BVW} in some more detail. In section 3 we present
the equations that constrain weak coupling type IIB flux vacua. In
section 4 it is shown that AdS type weak coupling solutions can be
found with masses of the order of the cosmological constant.
Minkowski space solutions are shown to live at strong coupling. In
section 5 we present a mechanism to break supersymmetry in terms of
fluxes and argue that the masses of states can in principle be much
larger than the cosmological constant once the theory is uplifted to
dS space. In section 6 we present our conclusions and open
questions.

\section{A Reminder of the Basics}

In this section we summarize some material concerning the type IIB
non-geometric model constructed in \cite{BVW} that will be relevant
in the later sections.

The concrete example we are interested in is a non-geometric model
described in terms of an orientifold a of Landau-Ginzburg (LG)
model. In \cite{BVW} it was shown that several types of
four-dimensional supersymmetric vacua (Minkowski as well as AdS)
exist at the Fermat point of this model. Even though these vacua
emerge at strong coupling their existence is warranted due to a
powerful non-renormalization theorem for the type IIB superpotential
\cite{Vafaaug}, \cite{DV}, \cite{Burgess}, \cite{BVW}. We will
discuss the non-renormalization theorem in some detail in this
section. In order to compute e.g. the explicit value of the masses
of moduli, it is of importance to find weak coupling solutions, as
will be done in the following sections.

\subsection{The model}
\label{model} We would like to compactify the type IIB theory on the
mirror of a rigid Calabi-Yau manifold \footnote{We restrict here to
the model with $h_{21}=84$ for concreteness, even though a model
with $h_{21}=90$ was also constructed in  \cite{BVW}. Our subsequent
analysis will be similar for the second model constructed in
\cite{BVW}.}with $h_{11}=0$ and $h_{21}=84$. The theory we are
interested in is constructed in terms of a LG model based on nine
minimal models and worldsheet superpotential
\begin{equation}
\eqlabel{fermat}
W=\sum_{i=1}^9 x_i^3,
\end{equation}
divided by a ${\zet}_3$ symmetry generated by
\begin{equation}
g:x_i \rightarrow \omega x_i,
\end{equation}
with $i=1,\dots,9$ and $\omega=e^{{2\pi i\over 3}}$. Having
$h_{11}=0$ the model has no K\"ahler moduli (or hypermultiplets
except for the universal hypermultiplet) so that it is intrinsically
non-geometric.

Constraints following from the tadpole cancellation condition can be
satisfied by considering an orientifold of this LG model which can
be obtained by dividing by worldsheet parity $\Omega_B$ dressed with
a holomorphic involution $P$ in space-time
\begin{equation}
P: (x_1,x_2, x_3 \dots x_9)=-(x_2,x_1,x_3,x_4,\dots , x_9).
\end{equation}
Complex structure deformations emerge as deformations of $W$ and a
basis of such deformations is given in terms of the invariant
monomials of the chiral ring
\begin{equation}
x_ix_jx_k,
\end{equation}
where $i\neq j\neq k\neq i$. There are $h_{21}=63$ monomials which
are invariant under the ${\zet}_3$ action as well as the orientifold
action. There are no K\"ahler structure deformations so that the
size of the manifold is not a modulus. Due to this fact, the
supergravity approximation is (strictly speaking) not valid and CFT
techniques need to be used to describe the internal theory.
Nevertheless the vacua that we will present are perturbatively and
non-perturbatively stable. As argued in \cite{BVW} this is due to
the existence of a non-renormalization theorem of the type IIB flux
superpotential, which will be presented in the next subsection in
some detail.

The above LG model can also be thought of as a toroidal orbifold,
where the orbifold group is a non-geometrically acting quantum
symmetry \cite{Vafaorbifold} \cite{BVW}. The type IIA mirror of this
model is however geometric. It can be represented in terms of a
torus $T^6=(T^2)^3$ divided by a ${\zet}_3 \times {\zet}_3$ symmetry
generated by
\begin{equation}
\eqlabel{generated}
\begin{split}
g_{12}&: (z_1,z_2,z_3) \rightarrow (\omega z_1,\omega^{-1} z_2,z_3),
\\
g_{23}&: (z_1,z_2,z_3) \rightarrow (z_1,\omega z_2,\omega^{-1}z_3),
\end{split}
\end{equation}
where $z_i$, $i=1,2,3$ are the coordinates of the torus.

{} From this representation it becomes obvious that the mirror of
our model is related to the toroidal model presented in \cite{wati},
though the action of the orbifold symmetry is somewhat different.
Similarly as in \cite{wati} our model has three bulk moduli that
correspond to the complex structures of the three tori, while the
remaining moduli correspond to blow up modes.

Here we will focus on stabilizing the bulk moduli and we shall
assume that blow up modes have been stabilized at a different scale.
This will already give us some useful information about the moduli
space. In the next section it is shown how the equations
constraining the bulk moduli of the toroidal orbifold of \cite{wati}
emerge as a special case of our equations.

\subsection{Non-renormalization theorem}

There are several arguments supporting the non-renormalization
theorem of the superpotential for complex structure moduli in the
type IIB theory. First indirect evidence came from the matrix model
calculation of \cite{DV}. Here we shall elaborate a more direct
argument based on supersymmetry that was already presented in
\cite{BVW}. In general, the superpotential of an $\caln=1$ theory
receives non-perturbative corrections. However, the particular
$\caln=1$ theory we are interested in has its origin in a theory
with $\caln=2$ supersymmetry, so that corrections to the
superpotential are absent.

Compactifying the type IIB theory on a Calabi-Yau threefold will in
general result in a theory with $\caln=2$ supersymmetry in $d=4$
containing $h^{21}$ vector multiplets and $h^{11}+1$ hypermultiplets
(where the `1' denotes the universal hypermultiplet). The $\caln=2$
supersymmetry can be broken to $\caln=1$ by adding a
Fayet-Iliopoulos term \cite{APT}, \cite{TV}, \cite{Ivanov}. We are
interested in the case $h^{11}=0$ which describes a theory with
$h^{21}$ abelian vector multiplets and the universal hypermultiplet.
To illustrate the argument it suffices to consider the simpler case
of a theory with only one abelian  $\caln=2$ vector multiplet, which
has an expansion in terms of $\caln=1$ superfields
\begin{equation}
\Phi=\phi^{(1)}(\tilde y,\theta)+{\sqrt 2} \tilde \theta^{\alpha}
\phi^{(2)}_{\alpha}(\tilde y,\theta)+\tilde \theta^{\alpha} \tilde
\theta_{\alpha} \phi ^{(3)}(\tilde y,\theta),
\end{equation}
where $\tilde y^{\mu}=x^{\mu}+i\theta \sigma^{\mu}\bar \theta +i
\tilde \theta \sigma ^{\mu} \bar{\tilde{\theta}}$, and $\theta$,
$\tilde \theta$ and their complex conjugates are superspace
coordinates. The action for the  $\caln=2$ vector then takes the
form
\begin{equation}
\int d^2\theta d^2\tilde \theta {\cal F}_0(\Phi),
\end{equation}
where ${\cal F}_0$ is the $\caln=2$ prepotential. Supersymmetry can
be broken by adding an FI term
\begin{equation}
\int d^2\theta d^2\tilde \theta {\cal F}_0(\Phi)+\xi D,
\end{equation}
where $D$ is one of the three auxiliary fields of the $\caln=2$
vector multiplet. Integrating out the auxiliary field exactly
reproduces the $\caln=1$ flux superpotential containing $H_{RR}$
\begin{equation}
W=\int H_{RR} \wedge \Omega,
\end{equation}
as $H_{RR}$ appears as an auxiliary field in the ${\cal N}=2$ vector
multiplet. The superpotential is not renormalized neither
perturbatively nor non-perturbatively as the $\caln=2$  prepotential
for vector multiplets is not renormalized \cite{Vafaaug}. One simple
argument for this is that in $\caln=2$ supersymmetric theories
neutral hypermultiplets and vector multiplets do not couple and the
dilaton is part of a hypermultiplet. Furthermore, $\alpha'$
corrections can be excluded as they contain the size of the
Calabi-Yau which is a K\"ahler modulus. Using the $SL(2,\zet)$
symmetry of the type IIB theory we can write down the unique
$SL(2,\zet)$ invariant combination of the superpotential
\begin{equation}
W=\int G \wedge \Omega,
\end{equation}
where $G=H_{RR}-\tau H_{NS}$. The complete superpotential is then
not renormalized as the $\caln=2$ prepotential is not corrected.
This argument is already enough to ensure the existence of Minkowski
vacua, which are solutions to the equation $\partial_i W=0$, where
`$i$' denotes the moduli. Some more thought is needed to guarantee
the existence of supersymmetric AdS vacua which are solutions to the
equation $D_iW=\partial_iW+\partial_iK W=0$. This is due to K\"ahler
invariance \cite{BVW}. Namely suppose we can choose the coordinates
for the moduli $t_i$ in such a way that the K\"ahler potential has
an expansion
\begin{equation}
K=t_i\bar t_i+a_{ij}t_i\bar t_j f(t,\bar t),
\end{equation}
where $t_i=\bar t_i=0$ describes our solution. Quantum corrections
to $\partial_iK$ evaluated at $t_i=\bar t_i=0$ will affect the
solution only by terms which are purely holomorphic
\begin{equation}
\delta K=\delta f(t)+cc.
\end{equation}
Such a correction can, however, be absorbed into the superpotential
which is a holomorphic section of a line bundle $W\rightarrow
exp(-\delta f(t))W$, so that the solution is unaffected.

The non-perturbative non-renormalization theorem of the type IIB
superpotential is valid for models containing no K\"ahler moduli.
The mirror type IIA interpretation of this result is that for models
with only one three-cycle there are no non-perturbative corrections,
as the only available three-cycle contains $H_{NS}$ flux. We will
return to possible non-perturbative corrections in Section 5.

\section{IIB Flux Vacua at Weak Coupling}

In this section we derive the equations constraining the three bulk
moduli. We shall see that mirror symmetry and the duality to the IIA
model of \cite{wati} predicts a subtle correction to the dilaton
K\"ahler potential in type IIB, which cannot be obtained by a
Kaluza-Klein reduction, in accord with the fact that our model is
not geometric in the first place. We will see that this correction
opens new directions in the landscape leading to an infinite set of
new AdS type vacua in the large complex structure limit! Minkowski
vacua with stabilized moduli are unaffected and constrained to live
at strong coupling \cite{BVW}.

\subsection{Deformations}

The first step in our construction is to go away from the Fermat
point \eqref{fermat}. More precisely, we shall be interested in the
LG model with superpotential
\begin{equation}
\eqlabel{deform}
W = \sum_{i=1}^9 x_i^3 + a_1 x_1x_2x_3 + a_2x_4x_5x_6+a_3 x_7x_8x_9
+ \cdots
\end{equation}
We intend to make $a_1$, $a_2$, $a_3$ large, and the dots are all
other possible cubic monomials which however we will assume to be
small compared with the $a_i$. The point of the deformation
\eqref{deform} is that in the limit of large $a_i$, the LG-model is
precisely mirror to the rigid $T^6$ orbifold \eqref{generated}. This
is clear from the standard LG description of the moduli space of
$T^2$. The remaining deformations of $W$ correspond to blowup moduli
from the $T^6$ perspective. By going to this region of moduli space,
we are able to focus on the stabilization of the three ``bulk
moduli'' $a_1$, $a_2$, $a_3$. (All other $T^6$ moduli are projected
out by the orbifold.) Thus we reduce our computation to one in the
neighborhood of three copies of the large complex structure limit of
$T^2$. The twisted sector moduli can then in principle be stabilized
by turning on fluxes through the blowup cycles. This can be
justified in principle by an exact computation of the periods, which
was done for the model at hand in \cite{cdp}, or alternatively along
the lines of \cite{wati}.

\subsection{Potentials for bulk moduli}

Let us denote the three (complex) bulk moduli as $t_1$, $t_2$,
$t_3$. These moduli fields (which are related to the $a_i$ in
\eqref{deform} in the usual way \cite{Candelas}, \cite{Lerche})
represent the three complex structures of $T^6=(T^2)^3$. At this
point we will write the formulas with all $t_i$'s in place and
consider the case $t_1=t_2=t_3$ later on for calculational
simplicity (though we would like to conjecture that our conclusions
also hold for the more generic situation).

Let us first describe the superpotentials and K\"ahler potentials
that are needed to describe the constraining equations for the bulk
moduli. The type IIB flux superpotential \footnote{\label{noted}
Note that the model is not geometric, so that the integrals that
follow have to be interpreted from the conformal fields theory point
of view, as was done in \cite{BVW}.}
\begin{equation}
W = \int (H_{RR} - \tau H_{NS}) \wedge\Omega = W_{RR}- \tau W_{NS},
\end{equation}
is described in terms of complex structure moduli and the dilaton,
while K\"ahler structure moduli do not appear. Here $\Omega$ denotes
the holomorphic three form of the internal space and
$\tau=C_0+ie^{-\phi}$ is the axio-dilaton combination. Let us
rewrite this superpotential in terms of the bulk moduli by expanding
the three form fluxes in terms of a cohomology basis dual to the
symplectic basis $(A^I,B_J)$ of  $H_3(M,\zet)$. We choose the
intersection numbers to satisfy
\begin{equation}
A^I \cap B_J=-B_J \cap A^I = \delta_J^I \qquad{\rm and } \qquad
A^I\cap A^J=B_I\cap B_J=0.
\end{equation}
The dual cohomology basis is denoted by $(\alpha_I, \beta^J)$ and
satisfies
\begin{equation}
\eqlabel{basis} \int_{A^J}\alpha_I=\int \alpha_I \wedge \beta^J =
\delta_I^J \qquad{\rm and } \qquad  \int_{B_J} \beta^I=\int \beta^I
\wedge \alpha_J=-\delta_J^I.
\end{equation}

In terms of this basis the fluxes can be expanded as
\begin{equation}
\eqlabel{fluxes}
\begin{split}
H_{RR}&=M_0 \a_0+M_2(\a_1+\a_2+\a_3)
-M_4(\b^1+\b^2+\b^3)-M_6\b^0,\\
H_{NS}&=N_0 \a_0+N_2(\a_1+\a_2+\a_3)
-N_4(\b^1+\b^2+\b^3)-N_6\b^0,\\
\end{split}
\end{equation}
where the $M_p$'s and $N_p$'s are the flux numbers. Furthermore we
take into account that the A-periods of $\Omega$ determine the
coordinates on moduli space, while the B-periods determine the
derivatives of the prepotential

\begin{equation} z^I=\int_{A^I} \Omega, \qquad {\cal G}_I(z) = \int_{B_I}
\Omega.
\end{equation}
In the large complex structure limit the prepotential of the model
takes the form \cite{Candelas}
\begin{equation}
{\cal G}(z)=-{\frac{1}{3!}} \kappa_{IJK} {\frac{z^I z^J z^K}{z^0}} ,
\end{equation}
where $\kappa_{IJK}$ are the Yukawa couplings. Since the model can
be related to a torus $T^6$ the only non-vanishing Yukawa coupling
is $\kappa_{123}=1$ (and symmetric permutations of the indices). The
resulting superpotential is
\begin{equation}
\begin{split}
W_{RR} &=  -t_1t_2t_3 M_0 + (t_1t_2+t_1t_3+t_2t_3) M_2 + (t_1+t_2+t_3)M_4 + M_6 ,\\
W_{NS} &=  - t_1t_2t_3 N_0 + (t_1t_2+t_1t_3+t_2t_3) N_2 + (t_1+t_2+t_3)N_4 + N_6, \\
\end{split}
\end{equation}
which is written in terms of the affine coordinates
$t^\alpha=z^\alpha/z^0$.

Having the form of the superpotential we need the form of the
K\"ahler potential for the axio-dilaton and bulk moduli. This
potential takes the form
\begin{equation}
\eqlabel{ourkahler} K =-\log\left[ \ii (t_1-\bar t_1) (t_2-\bar
t_2)(t_3-\bar t_3) (\tau-\bar\tau)^4\right].
\end{equation}
We need to pause to explain the structure of the K\"ahler potential
for the axio-dilaton, $-4\log(\tau-\bar\tau)$ as opposed to the
familiar $-\log(\tau-\bar\tau)$ (compare e.g. with the appendix of
\cite{Giddings}). Recall that one way to define the model is to
start on a large volume $T^6$, go to a symmetric point in K\"ahler
moduli space and then orbifold in such a way as to project out all
K\"ahler moduli (see Section \ref{model}). Let us rephrase this in
supergravity language, as this allows for an easier comparison with
type IIA analysis of \cite{wati}. Assume that we have a four
dimensional $\caln=2$ supergravity with some number $n_h$ of
hypermultiplets, one of which is the universal hypermultiplet and
some number $n_v$ of vector multiplets. In many cases, the moduli
space of hypermultiplets contains a region in which the prepotential
is cubic, and which can be derived by dimensional reduction from ten
dimensional supergravity. Let us call it the cubic region. In type
IIB this is simply the large volume regime. In type IIA, we have to
be in the large volume and also in the ``large complex structure
limit''.

We now break $\caln=2$  to $\caln=1$ by an orientifold action and
fluxes. After the orientifold projection we remain with $n_v^-$
$\caln=1$ vector multiplets and $n_h+n_v^+$ $\caln=1$ chiral
multiplets. As is customary, we will continue to call the $\caln=1$
chiral multiplets which come from $\caln=2$ vector multiplets as
vector multiplet moduli and those which come from $\caln=2$
hypermultiplets as hypermultiplet moduli.

Now it was shown in \cite{grilo}, \cite{grimm} that after
orientifold projection the K\"ahler potential for the
hypermultiplets in the cubic region is of the form
\begin{equation}
\eqlabel{gl} K = - \log(\tau-\bar\tau) - 2\log
\ee^{-3\phi/2}\kappa_{abc} v^av^bv^c,
\end{equation}
where $v^a$ are some real coordinates on the hypermultiplet moduli
space (K\"ahler moduli in type IIB) and $\kappa_{abc}$ are the
Yukawa couplings. See in particular chapter 3, section 3 of
\cite{grimm}. The main point of the discussion is that one should
view the $v^a$ as the worldsheet couplings, whereas the holomorphic
coordinates on the hypermultiplet moduli space from the spacetime
point of view contain as real part the ${\rm Re}(T_a) \sim
\ee^{-\phi/2}v^a$, where $\ee^{-\phi}= {\rm Im}(\tau)$ is the
dilaton. Therefore, in the large volume limit, the K\"ahler
potential for the dilaton indeed reduces to just
$-\log(\tau-\bar\tau)$, as expected from a Kaluza-Klein reduction.

Now on the torus the cubic expression \eqref{gl} is exact (as there
are no worldsheet instanton corrections). We can then go to the
$\zet_3\times \zet_3$ symmetric point $v^a=1/2$, and orbifold. Since
orbifolding is a worldsheet operation it projects the worldsheet
variables $v^a$ to their value at the orbifold point but does not
touch the dilaton directly. It then follows easily from \eqref{gl}
that after orientifold and orbifold the K\"ahler potential for the
dilaton is
\begin{equation}
K(\tau) = - 4\log(\tau-\bar\tau).
\end{equation}
In a certain sense, one should view the factor of $4$ as a small
volume correction to the usual expression and which emerges for the
type IIB non-geometric models. This correction can only be derived
by mirror symmetry and not by analytic continuation from a geometric
Kaluza-Klein reduction in type IIB! In fact, \cite{grilo,grimm}
give a more general expression for the K\"ahler potential valid (in
the supergravity approximation) throughout the hypermultiplet moduli
space and this is what is used in \cite{wati} in their general
analysis.

Notice that this {\it non-geometric} modification of the K\"ahler
potential only affects AdS type vacua, because the determining
equations for Minkowski vacua do not depend on the K\"ahler
potential.

\subsection{Tadpole cancellation and supersymmetry constraints}

The single most interesting aspect of the modified K\"ahler
potential for the axio-dilaton is that
unbroken supersymmetry no longer requires the three-form flux to be imaginary
self-dual (ISD) even in the absence of non-perturbative corrections.
This renders the flux contribution to the tadpole non-positive definite and
makes the tadpole cancellation condition less constraining than in the
usual cases. This is essentially what allows the existence of the
sequence of flux vacua with arbitrarily large flux numbers found in
\cite{wati}.

More concretely, using the explicit form of the flux superpotential
the supersymmetry constraint for the axio-dilaton reads \footnote{As
noted in the foot of page \pageref{noted}, the integrals here and
below obtain their meaning from the CFT/LG description of our
model.}
\begin{equation}
\eqlabel{dileom} D_\tau W =
-{1\over \tau - \bar \tau}\int (3 G + \bar G)\wedge \Omega = 0,
\end{equation}
which is solved by
\begin{equation}
\eqlabel{dilsolv} \tau = {1\over 2} {W_{RR}\over W_{NS}} \left( 3 -
e^{i \varphi}\right)\qquad {\rm where}\qquad \varphi={\rm Arg}\left(
W_{NS} \over W_{RR}\right).
\end{equation}
The supersymmetry constraint for the complex structure moduli is
expressed in terms of a basis $\chi_i$ of harmonic $(2,1)$ forms
\begin{equation}
D_i W = \int G\wedge \chi_i  = 0.
\end{equation}
{} From this we see that unbroken supersymmetry requires the Hodge
decomposition of $G$ to be
\begin{equation}
G = A^i\chi_i + A^0 (-3\Omega+\bar\Omega),
\end{equation}
where the $A$'s are constants. As a result $G$ can have a component
in the $(3,0)$ direction which is IASD, as opposed to the $(2,1)$ and $(0,3)$
components which are ISD. This violates the standard
lore according to which supersymmetric three-form fluxes in Calabi-Yau compactifications
of type IIB string theory are constrained to be $(2,1)$ and in particular ISD.
Moreover, this renders the flux contribution to the tadpole non-positive definite.
To see this we write the tadpole in the form
\begin{equation}
\int H_{RR} \wedge H_{NS} =\ii\ee^{\phi} \int G\wedge \bar G.
\end{equation}
Using that the metric on moduli space $g_{i \bar j}$
and the K\"ahler potential $K$ are represented by
\begin{equation}
g_{i \bar j}=\ii e^K \int\chi_i \wedge\bar\chi_j  > 0 \qquad {\rm and } \qquad
\ii\int\bar\Omega\wedge\Omega = e^K> 0,
\end{equation}
and are both positive definite we obtain a negative contribution to
the tadpole from a particular component of $G$ namely
\begin{equation}
i\int (-3\Omega + \bar\Omega) \wedge(-3\bar\Omega+\Omega) = - 8 e^K <
0.
\end{equation}
Thus by turning on enough of this flux component we get an
indefinite space of supersymmetric fluxes. This results in flux
vacua with a finite tadpole in new directions of flux space. Some of
these directions allow flux numbers tending to infinity. In the
following we would like to discuss the properties of the string
theory landscape along these directions.

\section{Supersymmetric Solutions}

Our goal in this section is to search for supersymmetric solutions
at weak coupling and for large values of the complex structure. We
shall look for solutions with diagonal complex structure
$(t_1=t_2=t_3)$ and will later conjecture that this simpler
situation reproduces all the features of the more generic situation.
For three equal complex structures the flux superpotentials take the
form
\begin{equation}
\eqlabel{diagonal}
\begin{split}
W_{RR}&= - t^3 M_0 + 3 t^2 M_2 + 3 t M_4 + M_6, \\
W_{NS}  &= - t^3 N_0 + 3 t^2 N_2 + 3 tN_4+N_6.
\end{split}
\end{equation}
The K\"ahler potential for the bulk moduli and axio-dilaton is
\begin{equation}
\eqlabel{kahler} K =-\log\left[  \ii (t-\bar t)^3 (\tau -
\bar\tau)^4\right].
\end{equation}
Using this form of the potentials we shall now look for Minkowski as
well as AdS type supersymmetric solutions.

\subsection{Minkowski space solutions}
In the following we will see that supersymmetric Minkowski space
solutions do not emerge for large complex structure but are confined
to finite value of the complex structure and strong coupling. These
are the solutions presented in \cite{BVW}. Groundstates
corresponding to a four-dimensional Minkowski space are obtained as
solutions of
\begin{equation}
\eqlabel{minkow1} W=D_\tau W=D_t W=0,
\end{equation}
which can equivalently be written as
\begin{equation}
\eqlabel{minkow2} W_{RR}=W_{NS}=0\qquad {\rm and } \qquad \tau =
{W'_{RR}(t)\over W'_{NS}(t)}.
\end{equation}
We are interested in finding physical solutions for which the
imaginary parts of $t$ and $\tau$ are non-vanishing, because
otherwise the solutions lie at the boundary of the moduli space.
Since vanishing of the superpotentials \eqref{diagonal} results in
cubic equations with real coefficients, complex solutions of these
equations only exist if the cubic polynomials $W_{RR}$ and $W_{NS}$
have two common complex conjugate roots, {\it i.e.} if they
factorize according to
\begin{equation}
\begin{split}
W_{RR}& =C(t-\alpha)(t-a)(t-\bar a), \\
W_{NS}&= D(t-\beta)(t-a)(t-\bar a), \\
\end{split}
\end{equation}
where $C,D,\alpha,\beta$ are real and $a$ is complex. These numbers
are constrained by flux quantization condition. The complex
structure is then determined from the zeros of a quadratic equation,
or equivalently
\begin{equation}\eqlabel{quads}
t = a=a_1+i a_2.
\end{equation}
Moreover, taking into account Eqs.
 (\ref{basis}), (\ref{fluxes}), (\ref{minkow1}), (\ref{minkow2}) the
tadpole can be written in the form
\begin{equation}
\eqlabel{tads} \int H_{RR} \wedge H_{NS}=4(M_2 N_0-N_2 M_0) (a_2)^2.
\end{equation}
Note that the coefficient in front of $(a_2)^2$ is integer and
cannot be made arbitrarily small nor equal to zero (the lhs of the
equation does never vanish in Minkowski). As a result $a_2$ is
bounded by the O3 plane charge. In the concrete examples considered
in \cite{BVW} the largest value of the RR charge arising from an O3
plane was 12. As a result Minkowski space solutions only exist if
the imaginary part of the complex structure is small like for
example in the solutions found in \cite{BVW}. In the type IIA mirror
these are solutions at small volume of the internal geometry, for
which no perturbative control is expected. In the following we will
see that AdS solutions do exist in the large complex structure
limit.

\subsection{Constraints for AdS type solutions}

Supersymmetric flux configurations which allow a negative
cosmological constant in the external space-time are the solutions
of
\begin{equation}
D_\tau W=D_t W=0.
\end{equation}
Since the vanishing of $W$ is no longer required, solutions of AdS
type are less constrained. It is useful to first rewrite the
superpotential in a more practical manner. For this take into
account that the equation $D_{\tau}W=0$ allows us to write $\tau$ in
the form\footnote{It is sensible to divide by $W_{RR}$ or $W_{NS}$
as Minkowski solutions are not a subclass of the AdS type solutions
derived in the following.}
\begin{equation}
\eqlabel{tau}
\tau = \frac 12 {W_{RR}\over W_{NS}}(3-e^{i \varphi}).
\end{equation}
Inserting this form of $\tau$ into the superpotential we get a
representation of the superpotential in terms of $W_{RR}$ up to a
phase
\begin{equation}
\label{superpot} W=\frac 12 W_{RR} (-1+e^{i \varphi}).
\end{equation}
Using Eqs. (\ref{tau}) and (\ref{superpot}) as well as the form of
the K\"ahler potential (\ref{kahler}) it is easy to see that the
constraint $D_t W=0$ takes the form
\begin{equation}\label{ci}
{\partial_t W_{RR}\over W_{RR}} -\frac 12 (3-e^{i \varphi})
{\partial_t  W_{NS}\over W_{NS}}=- \frac 32 \frac{1}{t-\bar t}
(1-e^{i \varphi}).
\end{equation}
This will be our starting point in the search for supersymmetric AdS
configurations.
\subsection{Weak coupling AdS type solutions}

Using the ansatz (\ref{diagonal}) for the superpotentials and the
constraints (\ref{tau}) and (\ref{ci}) we obtain the solution to the
supersymmetry constraints. Different types of solutions are
possible. Our aim is to present some concrete examples, leaving the
search for the most general solution for future work.

\subsubsection{Constant $W_{NS}$}

The simplest solution of Eq. (\ref{ci}) is given by flux
configurations in which $W_{NS}$ is constant, {\it i.e.},
\begin{equation}
W_{NS}=N_6\qquad {\rm while} \qquad  W_{RR}=-t^3 M_0 + 3 t^2 M_2 + 3
t M_4 + M_6,
\end{equation}
where the flux numbers $N_0$, $N_2$ and $N_4$ are set to zero. The
corresponding fluxes given by Eq. (\ref{fluxes}) induce a
contribution to the tadpole of the form
\begin{equation}
\int H_{RR} \wedge H_{NS}=-M_0 N_6,
\end{equation}
which only constrains the flux numbers $M_0$ and $N_6$.

The modulus $t=t_1 + i t_2$ is determined from Eq. (\ref{ci}) which
for real and constant $W_{NS}$ takes the form
\begin{equation}
\partial_t W_{RR} +\frac 32 {{\rm Im}(W_{RR}) \over {\rm Im} (t)} =0.
\end{equation}
The imaginary part of this equation determines the real part of the
complex structure
\begin{equation}
t_1={M_2 \over M_0},
\end{equation}
while the real part of the equation determines $t_2$
\begin{equation}
t_2 = \sqrt{\frac 53}\sqrt{-\frac{M_2^2}{M_0^2}-\frac{M_4}{M_0}}.
\end{equation}
The imaginary part of the axio-dilaton which follows from Eq.
(\ref{tau}) is
\begin{equation}
\tau_2 = - \frac {24}{5} \frac{M_0}{N_6} (t_2)^3.
\end{equation}
Since $M_2$, $M_4$ and $M_6$ are not bounded by the orientifold
charge they can be made arbitrarily large, giving us small
parameters controlling the expansion, the inverse of the flux
numbers. As a result solutions exist in the large complex structure
and weak coupling limit, as becomes evident from the above
expressions for $t_2$ and $\tau_2$. Further, the axion $\tau_1$ is
fixed by Eq. (\ref{tau}). This example describes the solution
discussed in \cite{wati} as can be seen by comparing to expressions
(4.6), (4.20)-(4.22) of that paper\footnote{The reader should not
get confused by the counting of moduli, which is 36 in \cite{wati}
and 63 for us. The precise statement is that a subclass of our
models is mirror to \cite{wati}. Both models have 3 bulk moduli
describing tori coordinates and only differ in the number of blow up
modes, which are not considered here in detail.}. We thus see that
the non-geometric type IIB model is mirror to the massive type IIA
model of \cite{wati} for a particular choice of our flux quantum
numbers! However in type IIB we have more freedom to dial the
fluxes, so that more general solutions can be constructed. Let us
see an example of this next.

\subsubsection{Non-constant $W_{NS}$}

Besides the solutions described in the previous paragraph there are
more directions in the landscape parametrized by different
combinations of flux numbers $N_i$ and $M_i$. One such example can
be constructed in terms of flux configurations with
\begin{equation}
W_{RR}=3 M_2 t^2 + M_6 \qquad {\rm and } \qquad W_{NS}=3 N_4 t.
\end{equation}
The contribution of the fluxes to the tadpole is
\begin{equation}
\int H_{RR} \wedge H_{NS} = 3 M_2 N_4,
\end{equation}
and as a result the O3 plane charge limits the values of $M_2$ and
$N_4$ while $M_6$ can be taken to be arbitrarily large. The
imaginary part of the complex structure and the axio-dilaton are
given by
\begin{equation}
\label{adsweak} t_2 = \sqrt{-\frac{M_6}{9 M_2}} \qquad {\rm and }
\qquad \tau_2= - 8 i \frac{M_2}{N_4} \sqrt{t_2},
\end{equation}
while the corresponding axionic partners (described in terms of the
real part of $t$) vanish. (But, as will be clear from the expressions
for the mass matrix in the next subsection, both $\tau_1$ and $t_1$
have a non-zero mass in those solutions.) From the result (\ref{adsweak}) 
we observe that by taking $M_2,N_4\sim O(1)$ and $M_6$ large we obtain 
a weak coupling solution in the large complex structure limit. The value
of the moduli scale in a different manner with the fluxes as for
the solutions in the previous subsection.

\subsection{The mass matrix}

\def\taubar{\bar{\tau}}
\def\tb{\bar{t}}

In the following we would like to show that in the large complex
structure limit a generic property of the mass matrix of the moduli
fields at the supersymmetric AdS groundstates is that it is of the
order of the space-time cosmological constant.

The masses of moduli fields are determined from the second
derivatives of the scalar potential at the groundstate which are
given by
\begin{equation}\eqlabel{hessian}
\begin{split}
\partial_{\bar b} \partial_a V & = e^K\left( D_a D_c W \overline{D_d D_a W}
g^{c \bar d} - 2 g_{a \bar b} \mid W \mid^2 \right),\\
\partial_a \partial_b V & = -e^K \left(D_a D_b W\right) \bar W. \\
\end{split}
\end{equation}
Here the indices $a,b,\dots$ label all the fields, {\it i.e.} the
complex structure and the axio-dilaton. To obtain these expressions
we have repeatedly used $D_a  W=0$ at the groundstate. For the
diagonal ansatz \eqref{diagonal}, we find the following expression
for the second K\"ahler derivatives of the superpotential at the
supersymmetric groundstate
\begin{equation}
\begin{split}
D_\tau D_\tau W &= -\frac{12 W}{(\tau-\taubar)^2},\\
D_t D_\tau W & = -D_t W_{NS}, \\
D_t D_t W &= -\frac{2(\tau-\taubar)}{t-\tb} \overline{D_t W_{NS}}.
\end{split}
\end{equation}
By introducing the parameters
\begin{equation}
\eqlabel{pars} x=(\tau-\taubar)(t-\tb)\frac{D_t W_{NS}}{W} \qquad
{\rm and } \qquad  y = (\tau-\taubar)(t-\tb)\frac{\overline{D_t
W_{NS}}}{W},
\end{equation}
the canonically normalized mass matrix for the moduli $t$, $\tau$
written as a hermitian matrix whose entries are given in Planck
units is
\begin{equation}
M_{\rm phys}^2/\Lambda_{AdS} =
\begin{pmatrix}
\frac 23 -\frac{10}{108} |x|^2 & \frac{2}{9} \bar y & -\frac{\bar x}{2\sqrt{3}}
-\frac{\bar y x}{9\sqrt{3}} & \frac{\bar x}{6 \sqrt{3}} \\
\frac 29y & \frac 23 - \frac{10}{108} |x|^2 & \frac{x}{6\sqrt{3}}
& -\frac x{2\sqrt{3}} -\frac{y \bar x}{9\sqrt{3}} \\
-\frac{x}{2\sqrt{3}} - \frac{y\bar x}{9\sqrt{3}} & \frac{\bar x}{6\sqrt{3}}
& -\frac 73-\frac{1}{36}|x|^2 & 1\\
\frac{x}{6\sqrt{3}} & -\frac{\bar x}{2\sqrt{3}} -\frac{x \bar y}{9\sqrt{3}}
& 1 & -\frac{7}{3}-\frac{1}{36} |x|^2
\end{pmatrix}.
\end{equation}
{} From this explicit expression we see that the essential parameter
that controls the masses is $x$ from \eqref{pars} (note that $y=\bar
x$ up to a phase). So we need to look for solutions of $D_t W=0$
with large $|x|$ and large $t_2$.

Next we use that $W_{RR}$ and $W_{NS}$ have one real and a pair of
complex conjugate roots. Since $W_{NS}$ is a polynomial of third
degree in $t$, having small $x$ is only possible if $t$ is close to
a zero of $W_{NS}$. Since moreover $t$ is a physical field, its
imaginary part should be non-vanishing. By Eq. (\ref{ci}) we also
need to be close to the zeros of $W_{RR}$. We write the
factorization of the cubic polynomials \eqref{diagonal} as
\begin{equation}
\eqlabel{factorize}
\begin{split}
W_{RR} &= C(t-\alpha)(t-a)(t-\bar a) \qquad \text{$\alpha$ real}\\
W_{NS} &= D(t-\beta)(t-b)(t-\bar b) \qquad \text{$\beta$ real}.
\end{split}
\end{equation}
Using this ansatz Eq. (\ref{ci}) reduces to
\begin{equation}\label{ask1}
\frac {1}{t-\alpha}+\frac {1}{t-a}+\frac {1}{t-\bar a}\approx\frac
12 (3-e^{i \varphi}) \left( \frac {1}{t-\beta}+\frac {1}{t-b}+\frac
{1}{t-\bar b}\right),
\end{equation}
where we have neglected the term on the right hand side of Eq.
(\ref{ci}) since it is subleading in the large complex structure
limit. In order to satisfy our constraints, the parameters $a$ and
$b$ need to have a large imaginary part, and we need to look for a
solution for which the real and imaginary parts of $t-a$, $t-b$ are
of order 1. It is not difficult to see that Eq. (\ref{ask1}) then
reduces to
\begin{equation}\label{ask2}
( t-b)\approx (t-a) A(t) ,
\end{equation}
where we have introduced the variable $A(t)$ related to the phase
$\varphi$ by
\begin{equation}
A(t) = \frac 12 (3-e^{i \varphi}).
\end{equation}
Note that $\varphi$ is approximately given by
\begin{equation}
e^{ i \varphi} \approx e^{ i \tilde \varphi} \left( {t-b \over \bar
t- \bar b}\right)\left( {\bar t-\bar a \over  t- a}\right) \qquad
{\rm where} \qquad e^{i \tilde \varphi } =\left( {{t-\alpha}\over
\bar t - \alpha}\right)\left( {\bar t - \beta \over t -
\beta}\right).
\end{equation}
The solvability of Eq. (\ref{ask2}) now depends on the value of the
phase $\tilde \varphi$. It is not difficult to see that if $e^{i
\tilde \varphi}$ is real Eq. (\ref{ask2}) has trivial solutions
only. On the other hand note that
\begin{equation}
e^{i \tilde \varphi}=\left(  t_1 - \alpha+ i t_2 \over t_1 - \alpha-
i t_2\right) \left(  t_1  - \beta- i t_2\over t_1 - \beta+ i t_2
\right).
\end{equation}
Since in the fundamental domain of $SL(2,\zet)$, $|t_1 |\leq 1/2$, a
complex phase can only be obtained in the large complex structure
limit if $\alpha$ and/or $\beta$ are proportional to $t_2$. However,
under this assumption the tadpole reduces to
\begin{equation}
\int H_{RR}\wedge H_{NS} = M_0 N_6-\frac 13 M_0N_0+\frac 13
M_2N_4-M_4N_2\approx C D \frac 23 (\alpha-\beta) {\rm Im}(a)^2 +
O({\rm Im} (a))
\end{equation}
which implies that $\alpha=\beta$ to leading order. This forces the
phase factor $e^{i \tilde \varphi}$ to be real.

We conclude that it is not possible to keep the masses large in such
a large flux limit in which the complex structure moduli become
very large while maintaining supersymmetry and the tadpole cancellation
condition. Notice that although we have not strictly imposed that
the dilaton also runs to infinity, this would not be an independent
constraint in our ansatz, in which $\tau\sim C/D$, see eq.\ \eqref{tau}.
We could have reached weak coupling if the tadpole had been
satisfied.

\section{Non-supersymmetric Solutions}

The supersymmetric AdS solutions of the previous section had moduli
fields with masses that were too small (of the order of the AdS scale)
so that these fields could be considered effectively as massless. We here
want to explore the possibility that this situation is remedied
after breaking supersymmetry by appropriate effects.

In a superficially similar situation, KKLT \cite{Kachru} proposed to
uplift type IIB AdS vacua to dS space by adding anti-D3 branes. The
resulting masses are of the order of the AdS scale {\it before}
the uplift so by appropriate fine-tuning will be large compared to
the positive cosmological constant after the uplift. Adding
anti-D3-branes is certainly one possibility to uplift our AdS type
vacua of Section 4 and to get masses of the right scale. However,
anti-D3-branes break supersymmetry by hand and we would like to
explore if a mechanism that breaks supersymmetry spontaneously (only
with fluxes) can be found. Such a mechanism was proposed in
\cite{Aganagic}, \cite{Heckman} for {\it non-compact} models. It
was shown there that a model with branes and anti-branes is
holographically dual to a model containing only fluxes after a
geometric transition. There is a simple generalization of
spontaneous supersymmetry breaking by fluxes for {\it compact}
models, that makes use of corrections to the scalar potential.

It was pointed out in \cite{vandoren} within the context of type IIA
compactifications on Calabi-Yau orientifolds that perturbative
corrections to the K\"ahler potential generate a contribution to the
potential for the dilaton which is similar to the contribution of
anti-D3 branes to the scalar potential for the radial modulus in
KKLT. Supersymmetry can be broken spontaneously instead of by
anti-branes. It was further argued in \cite{vandoren} that membrane
instantons \cite{Strominger1} generate a non-perturbative correction
to the superpotential of the dilaton, which resembles the
non-perturbative correction to the superpotential for the radial
modulus coming from gaugino condensation and wrapped D3 branes in
KKLT.

Mirror symmetry implies that higher order terms in the type IIB
theory result in perturbative corrections to the K\"ahler potential
for the dilaton\footnote{One of these terms was used in \cite{Louis}
to correct the K\"ahler potential, but it is possible to have more
contributions from the higher order terms in the action appearing in
\cite{Green}, \cite{Berkovits}.}. Membrane instantons correspond to
D(-1) instantons in the type IIB theory. The AdS type models
discussed in Section 4 contain no hypermultiplets (except for the
universal one), and do not have non-perturbative corrections to the
superpotential according to the non-renormalization theorem. From
the type IIA side this follows from the absence of membrane
instantons, as the only available three-cycle contains $H_{NS}$ flux
\cite{Strominger2}. Of course, more complicated models having more
than one three-cycle could receive non-perturbative corrections,
which will be discussed next or there could be corrections arising
from branes wrapping cycles or the whole Calabi-Yau which are
forbidden in supersymmetric solutions but which could appear once
supersymmetry is broken. It is quiet possible that perturbative
corrections to the K\"ahler potential lift the AdS type vacua of
Section 4 to dS space with moduli field masses of the right
magnitude. Once the numerical coefficients of all relevant type IIB
interactions are known, this can be checked explicitly.

It is curious that there is a second type of rather simple type IIB
vacua that break supersymmetry spontaneously. As we shall see these
vacua require no sinks of RR charge in order to satisfy the tadpole
cancellation condition on a compact Calabi-Yau three-fold since they
have $H_{NS}=0$\footnote{These supersymmetry breaking configurations
will be further discussed in \cite{stud}.}. Consider the scalar
potential of ${\cal N}=1$ supergravity in four dimensions
\begin{equation}
V=e^K \left( g^{a \bar b} D_a W \overline{ D_b W} - 3 |W|^2 \right),
\end{equation}
where in our case $a,b$ label the axio-dilaton and complex structure
moduli. It turns out that in this model moduli stabilization can be
achieved using RR three-form fluxes only, {\it i.e.} by assuming the
superpotential is of the form
\begin{equation}
W = W_{RR}  = \int H_{RR} \wedge \Omega.
\end{equation}
Since in this case $W$ is independent of $\tau$ it is a matter of
simple manipulations to show that
\begin{equation}\label{axxii}
V =e^K \left( g^{i \bar j} D_i W_{RR} \overline{D_j  W_{RR}} + |
W_{RR}|^2 \right),
\end{equation}
where $i,j$ label the complex structure moduli only. This expression
which closely resembles the scalar potential for non-supersymmetric
black holes is positive definite\footnote{The factor 4 in the
dilaton K\"ahler potential is important to achieve this.} so we can
hope to get vacua with a positive cosmological constant. However,
$\tau_2$ is not stabilized since the only dependence in $\tau$
appears in the overall factor $e^K \sim {\rm Im}(\tau)^{-4}$. This
last factor causes the dilaton to run to weak coupling.

One way to stabilize the axio-dilaton is to take non-perturbative
corrections to the superpotential into account. Whereas we have
argued in section 2 that non-perturbative effects do not affect our
supersymmetric solutions discussed before, one generically expects
such corrections in models with more than one hypermultiplet.
However, as pointed out above, if we turn on only $H_{RR}$ flux, we
can avoid orientifolding and then there are more possibilities for
non-perturbative corrections even in our models. (In type IIA
language, we have twice as many three-cycles to wrap membranes if we
do not orientifold.) The superpotential could then take the
schematic form
\begin{equation}\label{axxi}
W=W_{RR}+A e^{i a \tau}.
\end{equation}
where $A$ and $a$ are in principle functions of the vector multiplet
moduli. As a result of {\it pushing down} the previous vacua with
non-perturbative effects, AdS type vacua with a stabilized
axio-dilaton emerge! It is interesting to uplift these solutions to
dS, so that the masses of moduli fields become large enough. This
can again be done with perturbative corrections to the K\"ahler
potential.

\section{Conclusions and Open Questions}
One of the simplest models in which moduli stabilization can be
studied emerges when the type IIB theory is compactified on an
internal {\it non-geometric} theory for which K\"ahler moduli are
absent. Such a model was constructed in terms of a LG model in
\cite{BVW}. Since the flux superpotential for the type IIB theory
contains all the complex structure moduli and the dilaton, one could
anticipate that the moduli can be stabilized at the classical level,
{\it i.e.} in terms of flux only. The resulting theory is then
extremely well under control because the anonymousness connected
with non-perturbative effects coming from branes is lacking. In
\cite{BVW} it was shown that Minkowski as well AdS solutions at
strong coupling with all moduli stabilized by fluxes can be found.
Even though these vacua live at strong coupling, their existence is
guaranteed due to the non-renormalization of the type IIB flux
superpotential.

In order to calculate physically relevant quantities for these
vacua, such as the mass matrix, it is important to find weak
coupling solutions. In this paper we have shown that
four-dimensional Minkowski solutions live at strong coupling, so
that the only Minkowski solutions of our model are the ones
presented in \cite{BVW}. We have also shown that weak coupling
solutions (large value of the complex structure and small value for
the dilaton) of AdS type exist. For a particular choice of flux
configuration our model is mirror to the type IIA model of
\cite{wati}.

We have computed the mass matrix of the AdS type weak coupling
solutions and we have seen that the masses of the states are
proportional to the four-dimensional cosmological constant. Being of
the order of the cosmological constant, one might be worried that
this implies that the masses of moduli are too small. However we
anticipated that this situation would change once supersymmetry is
broken and the model is uplifted to four-dimensional dS space. We
have discussed that {\it perturbative} corrections to the K\"ahler
potential coming from higher order interactions in the type IIB
theory break supersymmetry spontaneously. Once the vacua are
uplifted to dS space the moduli become heavy.

We further illustrated the existence of a  simple supersymmetry
breaking solution for which $H_{NS}=0$, so that no constraints are
imposed by the tadpole. Non-perturbative corrections to the
superpotential (if they exist) could stabilize the dilaton, while
perturbative corrections to the K\"ahler potential could uplift
these vacua to dS space.

There are several rather interesting open questions. We have
considered the technically simpler case of a $T^6$ with three equal
complex structures.  It would be interesting to construct the
generic weak coupling solutions, {\it i.e.} the solution for three
arbitrary complex structures and to check if the masses of moduli
are of order of the cosmological constant. We have conjectured that
this is generically the case in all models where parametric control
can be achieved, so that this would constitute a check of our
conjecture.

Since our Minkowski vacua (as well as many other flux vacua in the
literature) live at strong coupling, an important open question is
to understand if there exists a weakly coupled dual gauge theory
describing these vacua. Some earlier attempts to find a dual gauge
theory for the vacua of \cite{wati} were made in \cite{Banks}, where
the precise form of this gauge theory was nevertheless not found. In
this regard, it might be interesting to explore if a connection to
the recently constructed four-dimensional Chern-Simons gauge theory
in AdS can be found \cite{Schwarz}, \cite{Gaiotto}.

As opposed to what happens in KKLT, where free parameters coming
from the number of branes and anti-branes are present, in our model
there are no free constants, as these are determined by the value of
the corrections to the K\"ahler potential and the superpotential. It
will be fascinating, though probably a hard job, to compute the
exact numerical value of these constants. This teaches us that in
string theory there are no free lunches!

\begin{acknowledgments}
We would especially like to thank Shamit Kachru and Cumrun Vafa for
many interesting discussions and comments on our manuscript. We
further would like to thank Tom Banks, Aaron Bergmann, Yu-Chieh
Chung, Michael Dine, Guangyu Guo, Simeon Hellerman, Joe Polchinski,
Giovanni Villadoro, Stefan Vandoren, Edward Witten and Xi Yin for
valuable discussions and communications. The work of K.B.\ was
supported in part by NSF grants PHY-0505757 and the University of
Texas A\&M. The work of M.B.\ was supported by NSF grants
PHY-0505757 and the University of Texas A\&M. The work of J.W.\ was
supported in part by the Roger Dashen Membership at the Institute
for Advanced Study and by the NSF under grant number PHY-0503584.
K.B.\ would like to thank the Institute for Advanced Study at
Princeton, the Galileo Galilei Institute for Theoretical Physics and
the CERN theory division for hospitality and financial support at
different stages of this work. M.B.\ would like to thank the Harvard
department of physics and the Institute for Advanced Study at
Princeton for hospitality and financial support while this work was
carried out.
\end{acknowledgments}


\end{document}